\documentclass{aa}
\usepackage[varg]{txfonts}
\usepackage{graphicx}
\usepackage{amsmath}
\usepackage{amssymb}
\usepackage{amscd}
\usepackage{hyperref}
\usepackage{natbib}
\usepackage{xfrac}
\usepackage{algorithm,algorithmic}
\usepackage{xtab,booktabs}
\usepackage{float}

\begin{document}

\title{Synthetic direct demodulation method and its applications in 
  \emph{Insight}-HXMT data analysis
  \thanks{
    Supported by National Natural Science Foundation of China (11403014 and 11373025).
  }
}
\author{Zhuoxi ~Huo\inst{1,4}
  \and Yang ~Zhang\inst{2,3,4}
}

\institute{Qian Xuesen Laboratory of Space Technology, China Academy of Space Technology, Beijing 100094, China\\
  \email{huozhuoxi@qxslab.cn}
  \and Max Planck Institute for Chemical Physics of Solids, 01187 Dresden, Germany\\
  \email{yzhang@cpfs.mpg.de}
  \and Leibniz Institute for Solid State and Materials Research, IFW Dresden, 01069 Dresden, Germany
  \and Tsinghua Center for Astrophysics, Department of Physics, Tsinghua University, Beijing 100084, China
}

\abstract
{}
{A modulation equation relates the observed data to the object where the observation is approximated by a linear system.
  Reconstructing the object from the observed data is therefore is equivalent to solving the modulation equation.
  In this work we present the synthetic direct demodulation (synDD) method to reduce the dimensionality of a general modulation
  equation and solve the equation in its sparse representation.}
{A principal component analysis is used to reduce the dimensionality of the kernel matrix and
  \emph{k}-means clustering is applied to its sparse representation in order to decompose the kernel matrix
  into a weighted sum of a series of circulant matrices.
  The matrix-vector and matrix-matrix multiplication complexities are therefore reduced from polynomial time
  to linear-logarithmic time.
  A general statistical solution of the modulation equation in sparse representation is derived.
  Several data-analysis pipelines are designed for the Hard X-ray modulation Telescope (\emph{Insight}-HXMT) based on
  the synDD method.}
{In this approach, a large set of data originating from the same object but sampled irregularly and/or observed with
  different instruments in multiple epochs can be reduced simultaneously in a synthetic observation model.
  We suggest using the proposed synDD method in \emph{Insight}-HXMT data analysis especially for the detection of X-ray transients
  and monitoring  time-varying objects with scanning observations.}
{}


\titlerunning{synDD method and its applications in \emph{Insight}-HXMT data analysis}
\authorrunning{Z. ~Huo \& Y. ~Zhang}

\maketitle

\section{Introduction}
\label{sec:intro}
The newly launched Hard X-ray Modulation Telescope (\emph{Insight}-HXMT) is China's first X-ray astronomical satellite.
\emph{Insight}-HXMT carries three main payloads onboard: the High Energy telescope (HE), the Medium Energy telescope (ME) and the Low Energy telescope (LE) \citep{zhang2014}.
One of its primary scientific objectives is scanning the Galactic plane to monitor time-varying objects and to discover X-ray transients \citep{zhang2014}.
\emph{Insight}-HXMT is based on the direct demodulation (DD) method because all telescopes onboard are position-insensitive collimated detectors and images can only be reconstructed from the observed data by offline data analysis \citep{li1993dd,li1994dd,li2007,zhang2009}.

In the analysis of astronomical data, observed data can be modelled as object functions modulated by kernel functions, which characterise the observation process, mainly the instrument response.
For example, an observed image is a spatial distribution of objects modulated by an imaging system, where a point spread function (PSF) serves as the modulation
kernel function.
Observed spectra are object spectra modulated by energy response matrices.
The relations among the observed data, the object, and the observation process can be formulated with a Fredholm integral equation of the first kind:
\begin{equation}
  d\left(\boldsymbol{\omega}\right) = \int h\left(\boldsymbol{x}, \boldsymbol{\omega}\right) f\left(\boldsymbol{x}\right) \mathrm{d} \boldsymbol{x}\text{,}
  \label{equ:fredholm}
\end{equation}
where $d\left(\boldsymbol{\omega}\right)$, $h\left(\boldsymbol{x}, \boldsymbol{\omega}\right)$ and $f\left(\boldsymbol{x}\right)$ represent models of the observed data, the kernel, and the object, respectively, and $\boldsymbol{\omega}$ and $\boldsymbol{x}$ are coordinate variables for the observed data domain (e.g. pixel indices, energy channels, possible state parameters of instruments and so on) and the object domain (e.g. WCS celestial sphere coordinates), respectively.
The object model $f\left(\boldsymbol{x}\right)$ represents both the sources and the sky background.
The kernel $h\left(\boldsymbol{x}, \boldsymbol{\omega}\right)$ characterises the observation process including the instrument response as well as non-sky backgrounds; for example, dark currents, or cosmic rays for high-energy detectors. In order to solve the equation, both periodic calibration of instrument response and non-sky background estimation are necessary to estimate the kernel.
Differences between the true kernel and its estimation lead to systematic errors in the solution, that is, the estimated object.
Discussions of instrument response calibration, non-sky background modelling and estimation, and systematic error due to above issues are beyond the scope of this work.

The DD method is introduced by \citet{li1993dd} to solve the modulation equation and then to reconstruct the unknown object from the observed data.
In Eq. \ref{equ:fredholm} a model of the observed data instead of the data itself is given.
Since the data itself is a random outcome of the observation process, statistical solution of the modulation equation is achieved via the DD method according to the probability distribution of the observed data implied by its model.
For observed data dominated by Poisson fluctuation, the simplest implementation of the DD method degenerates into Richardson-Lucy (RL) iteration \citep{richardson1972,lucy1974}.
As a non-parametric approach, reconstruction through RL iteration allows a larger pool of solutions that fit the observed data, compared with parametric
fitting approaches that require explicit and known models of both signals and backgrounds \citep{puetter2005}.
However, a larger pool of solutions often results in an ill-posed problem, which is unstable or divergent.
The DD method overcomes this problem by continuously cross-correlating both sides of Eq. \ref{equ:fredholm} and transforming the original modulation equation into its $L$-fold equivalents to improve its positive definitiveness, as well as non-linear physical constraints used to regularize the iteration, thereby shrinking the pool of solutions so that the object reconstruction problem is reduced.

In real-world data analysis, it is often necessary to avoid data with a poor signal-to-noise ratio (S/N) through screening and selection; for example, the \emph{Good Time Interval} (GTI) auxiliary record \citep{blackburn1995ftools} is used to select time series data in \emph{Rossi X-ray Timing Explorer} (RXTE) as well as \emph{Insight}-HXMT data analysis, or, to include relevant data from other instruments or epochs in order to provide better statistics.
Meanwhile, the dimensionality and irregularity of the corresponding modulation equation increases, which imposes growing complexity on the reconstruction, especially for \emph{Insight}-HXMT data analysis, where the modulation kernels lack circular symmetry.
The computational complexity introduced by rotation modulation is reduced by an angular clustering method specifically designed for \emph{Insight}-HXMT all-sky survey \citep{huo2013raa}, since the rotation modulation can be represented much more sparsely with explicit angular coordinates.

Although the DD method treats the ill-posed object reconstruction problem, the lack of a modulation-kernel-constructing process or accelerated implementation for general cases prevents this method from being applied to data analysis tasks with high dimensionality or large datasets.
In this article we provide the synthetic direct demodulation method (synDD), which features
  \begin{enumerate}
    \item a modulation-kernel-constructing process to combine kernels characterizing individual instruments, observations and/or data screening/selection into one synthetic kernel (modulation equation synthesis process), and
    \item an accelerated implementation for general cases by decomposing an arbitrary kernel matrix into a weighted sum of a series of circulant matrices so that the fast Fourier transform (FFT) can help (modulation kernel matrix analysis process).
  \end{enumerate}
Consequently, we can not only deal with the previously mentioned rotation modulation without losing important information sampled from different position angles, as well as the projection distortion that occurs on the tangential plane of the celestial sphere, which we put aside in previous work \citep{huo2013raa}, but can also squeeze a more complicated kernel (e.g. with screening or weighing matrices) and additional observed data into our representation of the modulation equation.

\section{Method}
\subsection{Modulation equation}
In this article, vectors and matrices are highlighted in bold, while regular type face is used for scalars, plain sequences (or, 1D array), and multiple dimensional arrays.
Operations on plain arrays are element-wise if not specifically indicated.
For instance, $\vec{x}$ indicates a vector of $x_1,x_2,\dots,x_n$, while $x$ simply suggests a plain sequence of these elements.
Modular arithmetic is used for subscripts in this article.

Consider the algebraic form of Eq. \ref{equ:fredholm}
\begin{equation}
\vec{d} = \vec{H}\vec{f}
\label{equ:mod}
,\end{equation}
where $N \times 1$ column vector $\vec{d}$ and $M \times 1$ vector $\vec{f}$ are discrete samplings of the models of the observed data and the object, while $N \times M$ matrix $\vec{H}$ is the modulation kernel matrix, which is a discrete sampling of the modulation kernel.

\subsection{Analysis and synthesis of modulation equation}
Both $N$ and $M$ in Eq. \ref{equ:mod} are used to indicate the size of the object reconstruction problem.
Time costs in solving the modulation equation increase rapidly as the size of the problem increases unless the naive matrix-matrix and matrix-vector multiplications are carefully treated.
For $N \times N$ matrices and $N \times 1$ vectors the time costs of matrix-matrix and matrix-vector multiplications are proportional to $N^3$ and $N^2$ , respectively.
In computer science, such costs are measured with computational complexities, denoted by $O(N^3)$ (cubic complexity) and $O(N^2)$ (quadratic complexity).
If the complexity could be reduced from polynomials (cubic or quadratic) to quasilinear, for example, $O(N \log N)$ by replacing all the naive matrix mulplications with FFTs, time costs for a typical $512 \times 512$ \emph{Insight}-HXMT image reconstruct ($N = 512 \times 512$) can be reduced from $10^5$ seconds to a few seconds \citep{huo2013raa}.
In order to achieve such a  treatment, we decompose the kernel matrix $\vec{H}$, for this purpose taken to be an arbitrary matrix, into the sum of a finite sequence of circulant kernel matrices multiplied by diagonal coefficient matrices, as
\begin{equation}
\vec{H} = \sum_k \vec{A}_k \vec{H}_k \text{,}
\label{equ:decompose}
\end{equation}
where $\vec{A}_k$ is an $N \times N$ diagonal matrix serving as the coefficient of the $k$-th $N \times M$ kernel matrix $\vec{H}_k$, which is row-circulant.
Diagonal entries of $\vec{A}_k$ have values of either $0$ or $1$ only.
A matrix $\vec{H}_k$ is row-circulant if each of its row vectors is circular-shifted by one element to the right relative to its preceding row; that is, $H_{k,i,j}=H_{k,i+1,j+1}=h_{k,j-i+1}$ for all $i$ and $j$, where $\vec{h}_k$ is the first row vector of $\vec{H}_k$.
A row-circulant matrix is not necessarily square.
If the number of its rows is greater than the number of its columns ($N > M$), its excess rows are considered as circularly appended to the first $M$ rows of the matrix, while if $N < M$ the matrix is considered as the first $N$ rows that truncated from a $M \times M$ square circulant matrix.
In addition, it is required that
\begin{equation}
\vec{A}_k \vec{A_l} = \begin{cases}
  \vec{A}_k\text{,} & \text{if }k=l\\
  \vec{0}\text{,} & \text{otherwise}
\end{cases}\text{,}
\end{equation}
and
\begin{equation}
\sum_k \vec{A}_k = \vec{I}_N\text{,}
\end{equation}
where $\vec{I}_N$ is the $N \times N$ identity matrix.

Allowing $\vec{a}_k$ to be a column vector of diagonal elements of $\vec{A}_k$,
 Eq. \ref{equ:mod} becomes
\begin{equation}
  \vec{d} = \sum_k \vec{a}_k \cdot \left( \vec{f} \ast \vec{h}_k^{\star} \right)\text{.}
\label{equ:syn-mod}
\end{equation}
The operator $\cdot$ denotes the element-wise multiplication, the operator $\ast$ denotes the circular convolution, and $\vec{h}_k^{\star}$ is the reverse of $\vec{h}_k$, that is, $h_{k,i}^{\star}=h_{k,M+1-i}$.
We note that $\vec{a}_k$ has $N$ elements while $\vec{f} \ast \vec{h}_{k}^{\star}$ has $M$ elements so before calculating their element-wise multiplication the right operand $\vec{f} \ast \vec{h}_{k}^{\star}$ is truncated (if $N>M$) or padded (if $N<M$) to the same size as the left operand $\vec{a}_k$.
Such truncation or padding is applied when necessary in this article.

Decomposition of $\vec{d}$ is naturally derived as
\begin{equation}
  \vec{d} = \sum_k \vec{a}_k \cdot \vec{d}_k \text{,}
\label{equ:syn-obs}
\end{equation}
where the $k$-th element of the observed data $\vec{d}_k = \vec{f} \ast \vec{h}_k^{\star}$.

The term \emph{analysis} refers to separating the original modulation equation into a finite sequence of equations by decomposing the kernel matrix as well as the observed data, while the term \emph{synthesis} refers to combining a finite sequence of modulation equations to form an equivalent one.
Analysis of a modulation equation is useful for reducing the computational complexity of the corresponding inverse problem.
Data obtained from different observations of the same object can be synthesised as in Eq. \ref{equ:syn-obs} and demodulated as in Eq. \ref{equ:mod} or Eq. \ref{equ:syn-mod}.

\subsection{$L$-fold correlation}
The one-fold correlation of a modulation equation such as Eq. \ref{equ:mod} is achieved by left-multiplying both sides of the equation by the transpose of the kernel matrix, as
\begin{equation}
  \vec{H}^\mathrm{T}\vec{d} = \vec{H}^\mathrm{T}\vec{H}\vec{f}\text{.}
\end{equation}
In the correlated equation, the unknown object image remains $\vec{f}$ while the observed data as well as the kernel matrix are both transformed into one-fold correlated data $\vec{c}_1 = \vec{H}^\mathrm{T}\vec{d}$ and kernel $\vec{P}_1 = \vec{H}^\mathrm{T}\vec{H}$.
The $L$-fold correlation is achieved recursively as
\begin{equation}
\left\{
\begin{aligned}
\vec{P}_L &= \vec{P}^\mathrm{T}_{L-1}\vec{P}_{L-1}\\
\vec{c}_L &= \vec{P}^\mathrm{T}_{L-1}\vec{c}_{L-1}
\end{aligned}\right.\text{,}\forall L\geq 1\text{,}
\end{equation}
provided that $\vec{c}_0=\vec{d}$ and $\vec{P}_0=\vec{H}$ \citep{li1994dd}, so the $L$-fold correlated equation is
\begin{equation}
  \vec{c}_L = \vec{P}_L \vec{f}\text{.}
\end{equation}

Although $L$-fold correlation is preferred when using the DD method, it is difficult to compute with naive matrix multiplication when the kernel matrix is very large; for example, a $2^{20} \times 2^{20}$ kernel matrix, since the computational complexity of multiplication between two $N \times N$ matrices is $O(N^3)$ if performed naively.
With a more complicated algorithm the computational complexity can be reduced to $O(N^{2.376})$ \citep{coppersmith1990}, however it would still take hundreds of hours to compute the multiplication between two $2^{20} \times 2^{20}$ matrices.

We have derived a simple but efficient approach to compute the matrix multiplication from Eq. \ref{equ:decompose}.
First we compute the first column vector of $\vec{H}^\mathrm{T}\vec{H}$, as
\begin{equation}
\begin{split}
\vec{H}^\mathrm{T}\vec{H}\begin{pmatrix}
  1\\
  0\\
  \vdots\\
  0
  \end{pmatrix}&= \sum_{k,l}{\vec{H}_l}^\mathrm{T}{\vec{A}_l}^\mathrm{T}
  \vec{A}_k\vec{H}_k\begin{pmatrix}
  1\\
  0\\
  \vdots\\
  0
  \end{pmatrix} \\
  &= \sum_k {\vec{H}_k}^\mathrm{T}\vec{A}_k\vec{H}_k\begin{pmatrix}
  1\\
  0\\
  \vdots\\
  0
  \end{pmatrix} \\
  &= \sum_k h_k \ast \left(a_k   h_k^{\star}\right)
\end{split}\text{.}
\end{equation}
We can compute the remaining column vectors in the same way.
The time cost of computing each column vector is proportional to $K N \log N$, where $K$ is the number of circulant matrices, if the circulant convolution is calculated with the FFT algorithm.
For rotation modulation, $K$ is the number of position angle clusters, for wide-field image reconstruction, $K$ is the number of field-of-view (FoV) clusters, and for synthetic kernel, $K$ is the number of observations.
Because both the position angle and the FoV distortion vary gradually in practical observations, their clusters are always far less than pixels or bins of the observed data samples.
We found that $20$ to $100$ clusters are sufficient to approximate the kernel variation where the difference between the simulated true kernel and the kernel approximated with the weighted sum of circulant matrices are negligible compared to the required kernel calibration accuracy.
Therefore $K$ is always less than $N$ by several orders of magnitude and is independent of $N$. As a result, the complexity is reduced to $O(N^2 \log N)$.

\subsection{Screening and weighing}
We can avoid data with poor S/N (screening) or adjust their weights accordingly by introducing a weight matrix $\vec{M}$ into Eq. \ref{equ:mod},
\begin{equation}
  \vec{M}\vec{d} = \vec{M}\vec{H}\vec{f}\text{,}
  \label{equ:wgt-mod}
\end{equation}
where the weight matrix $\vec{M}$ is an $N \times N$ diagonal matrix.
The $i$-th element on its main diagonal $m_{i}$ serves as the weight of the $i$-th observed datum $d_i$.
We can avoid certain observed data by assigning $0$-valued elements as their weights.

Modulated estimates of the object $\vec{H}\vec{f}'$ ($\vec{f}'$ is an estimate of the unknown object $\vec{f}$) appear as denominators in reconstructions with the DD method as well as RL iterations.
The original kernel in Eq. \ref{equ:fredholm} contains both instrument response and non-sky background, while the object contains sources and the sky background; hence \emph{division-by-zero} will not happen.
But possible zero-weights in $\vec{M}$ of Eq. \ref{equ:wgt-mod} are like holes and may cause such a problem.
To prevent this problem in later reconstructions, $L$-fold correlation is applied so that \emph{zero-holes} in $\vec{M}\vec{H}\vec{f}'$ are \emph{smoothed} to non-zeros by the kernel and their non-zero neighbours.

\subsection{Additional observation}
\label{sec:multiple-epoch}
Observations at multiple epochs or with different instruments can be joined into a single modulation equation by using a partitioned kernel matrix, as
\begin{equation}
  \begin{pmatrix}
    \vec{d}_1 \\
    \vec{d}_2 \\
    \vdots \\
    \vec{d}_\Lambda
  \end{pmatrix} = \begin{pmatrix}
    \vec{H}_1 \\
    \vec{H}_2 \\
    \vdots \\
    \vec{H}_\Lambda
  \end{pmatrix} \vec{f}\text{,}
  \label{equ:block-mod}
\end{equation}
where $\vec{d}_\lambda$ and $\vec{H}_\lambda$ ($\lambda=1,2,\dots,\Lambda$) are the observed data and the kernel matrix of the $\lambda$-th observation, respectively.

\subsection{Iterative solution of synthetic modulation equation}
A model of the observed data instead of the observed data themselves has been placed on the left side of Eq. \ref{equ:fredholm} and Eq. \ref{equ:syn-mod}, where the model characterizes the statistically expected value of the observed data while the data itself is a random outcome from a specific observation.
For Poisson fluctuation(photon noise)-dominant data, the likelihood function of the object image $\vec{f}$ as an unknown parameter given the observed data $\vec{d}$ is
\begin{equation}
  \mathcal{L}(\vec{f}|\vec{d}) = \prod_{j}\dfrac{(\sum_{k,i}a_{k,j}f_j h_{k,i-j+1})^{d_j}e^{-\sum_{k,i} a_{k,j} f_j h_{k,i-j+1}}}{d_j !}\text{.}
\end{equation}

Once we find the $\vec{f}$ that maximizes the likelihood function, the maximum likelihood (ML) solution of the modulation equation is achieved.
The object reconstruction problem is therefore transformed into solving the following equation.
\begin{equation}
\frac{\partial \ln \mathcal{L}(\vec{f}|\vec{d})}{\partial f_i} = 0
  \text{,}\;\forall i\text{.}
\end{equation}
Therefore,
\begin{equation}
 \begin{split}
   &\sum_j \left( \dfrac{d_j}{\sum_{k,i}a_{k,j}f_i h_{k,i-j+1}} - 1\right)
     \sum_{k}a_{k,j}h_{k,i-j+1} \\
 = &\sum_{k,j} \dfrac{a_{k,j}h_{k,i-j+1}d_j}
  {\sum_{k,i}a_{k,j}f_i h_{k,j-i+1}^{\star}} -
  \sum_{k,j}a_{k,j}h_{k,i-j+1}
 = 0\text{,}\; \forall i \text{,}
 \end{split}
\end{equation}
and
\begin{equation}
\dfrac{\vec{f}}{\sum_{k}\vec{a}_k \ast \vec{h}_k}\sum_k
  \dfrac{\vec{a}_k \cdot \vec{d}}
  {\sum_k \vec{a}_k \cdot \left(\vec{f} \ast \vec{h}_k^{\star}\right)} \ast \vec{h}_k = \vec{f}
  \text{.}
\end{equation}

Fixed-point iteration is a method of finding a fixed point of a given function in a numerical analysis.
$x$ is a fixed point of the function $g(x)$ if $x = g(x)$.
The iteration $x^{(n+1)} = g\left(x^{(n)}\right)$ is expected to converge to $x$ if $g$ is continuous.
The ML estimate of the true image $f$ is a fixed point of function
\begin{equation}
\varphi(\vec{f}) = \dfrac{\vec{f}}{\sum_{k}\vec{a}_k \ast \vec{h}_k}\sum_k
  \dfrac{\vec{a}_k \cdot \vec{d}}
  {\sum_k \vec{a}_k \left(\vec{f} \ast \vec{h}_k^{\star}\right)} \ast \vec{h}_k \text{.}
\end{equation}
We therefore expect to find the ML estimate iteratively. Iteration at the $l$-th step is
\begin{equation}
  \vec{f}^{\left(l\right)} = \frac{\vec{f}^{\left(l-1\right)}}{\vec{w}} \sum_k\frac{\vec{a}_k \cdot \vec{d}}{\sum_k\vec{a}_k\cdot\left(\vec{f}^{\left(l-1\right)} \ast \vec{h}_k^{\star}\right)} \ast \vec{h}_k\text{,}
\label{equ:syndd}
\end{equation}
where the normalization factor $\vec{w} = \sum_k \vec{a}_k \ast \vec{h}_k$.
RL iteration can also be considered as a fixed-point iteration that achieves an ML solution that restores an image blurred by a convolution kernel.
For a perfectly known modulation kernel and statistical model of the observed data, the proper criterion to terminate the fixed-point iteration can be found by monitoring the residuals.
However when the modulation kernel is not perfectly known or the statistical model of the observed data does not perfectly describe the random nature of the data, one should terminate the iteration as soon as the required source is resolved from the data.
In addition, cross-validation and sensitivity assessment \citep{huo2015} are necessary to prevent overfitting or noise amplification.
For \emph{Insight}-HXMT survey data analysis, which is mainly focused on point-like-source detection and monitoring, the stopping criterion is not critical since the demodulated image is not the final result of object reconstruction but only serves as a hint for the following parametric fitting procedures.

Provided both the object image $f$ and the observed data $d$ are $N \times 1$ and the kernel matrix is then $N \times N$, it takes $\sim 2 N^2$ scalar multiplications for each RL iteration of the original DD.
In contrast, it takes $\sim 2 K   N \log N$ scalar multiplications for each iteration of demodulation synthesised from a  $K$ circulant kernel matrix with the aid of the FFT algorithm.
The computational complexity is therefore reduced, provided that $K \ll N$.

Let us take \emph{Insight}-HXMT as an example, where the level-1 scientific data products consist of lists of X-ray photon arrival events detected by each scientific instrument.
Each event is described by properties such as time on arrival, detector identifier, energy channel identifier, anti-coincidence detector counts, and so on.
Satellite orbit coordinates and telescope attitude (pointing angles and position angle) on arrival of each event are interpolated from the housekeeping data provided by the satellite platform.
Two-dimensional discrete samples of the observed image, 1D binned samples of the observed light curve, and/or a 1D binned energy spectrum are counted from the above events as well as auxiliary housekeeping data accordingly.
For object image reconstruction, each image contains too many pixels for the naive matrix multiplication in Eq. \ref{equ:mod} and the modulation cannot be computed with spherical harmonics or fourier transforms due to the absence of circular symmetry in the kernel, i.e. the point spread function (PSF) \citep{huo2013raa}.
With the approach described here the time cost of \emph{Insight}-HXMT image reconstruction is reduced by orders of magnitude.

\subsection{Cluster analysis of the modulation kernel}
\label{sec:cluster}
\emph{Cluster analysis} of the modulation kernel is the key building block of the method we present in this article.
A kernel matrix is an ordered set of row vectors, which are classified into $K$ groups through cluster analysis.
In cluster analysis we refer to each group as a cluster.
Row vectors in the same cluster are similar to each other, as if they were taken from a circulant matrix.
A generalized Euclidean distance $\lVert\vec{x}_i - \vec{x}_j\rVert$  is introduced here to measure the similarity between a pair of row vectors $\vec{x}_i$ and $\vec{x}_j$.
The two row vectors are considered sufficiently similar if the distance is less than the kernel calibration accuracy.
So cluster analysis is the underlying procedure through which a kernel matrix $\vec{H}$ is decomposed into a sequence of circulant matrices $\vec{H}_k$ and their coefficients $\vec{A}_k$.
The number of clusters, $K$, determines the extent to which the computational complexity can be reduced.
The smaller the number $K$, the more complexity is reduced when we approximate $\vec{H}$ with $\sum_{k=1}^K \vec{A}_k \vec{H}_k$.
The accuracy of the approximation is determined by similarities between any two row vectors in the same cluster.
Therefore, to perform a cluster analysis of a modulation kernel is to classify all row vectors of the kernel matrix into as few clusters as possible, while the similarities between any two vectors in the same cluster are acceptable.

\emph{Principal component analysis} (PCA) is also used to decompose a kernel in order to reduce its dimensionality \citep{jolliffe1986pca}.
We categorize the analysis of a modulation kernel in this article as a  cluster analysis.
Because PCA is only used to reduce the dimensionality of calculating the similarities between row vectors here, this means that instead of calculating the generalized Euclidean distance between two $M \times 1$ row vectors, we actually calculate the distance between two representative vectors that each contain many fewer components, i.e., the principal components.
Therefore, PCA serves as a preprocessing for cluster analysis.
Since each row vector of the kernel matrix has the same size as the object image, expense of computing similarities between row vectors increases sharply with the image resolution, especially if the image has more than one dimension.
Fortunately a row vector of a kernel could be more sparse with a certain representation than it appears with a naive pixel-wise representation.
PCA is used to find the basis of a sparse representation of given row vectors of a specific kernel.
In PCA the basis vectors are termed \emph{principal components}.
By expressing a row vector as a finite linear combination of the basis vectors, its dimensionality is effectively reduced, since the number of non-zero coefficients of a sparse representation is usually less than the number of pixels of the image.

Each row of the given kernel matrix $\vec{H}$ is circularly shifted leftwards according to its row number in the matrix, that is, the $i$-th row is shifted by $i$ elements leftwards.
In this way all row vectors of a circulant matrix would be aligned so that the shifted matrix appears as $N$ vertically stacked copies of identical row vectors, which is ready for the following PCA processing.

Allowing $\vec{x}_i$ be the shifted $i$-th row vector with pixel-wise representation, we use the NIPALS-PCA algorithm \citep{geladi1986} to find the principal components of the set of shifted row vectors $\{\vec{x}_i\}$ iteratively, and transform each row vector $\vec{x}_i$ to a new vector $\vec{t}_i$ in a space with reduced dimensionality defined by the principal components.

The similarity between two row vectors $\vec{x}_i$ and $\vec{x}_j$ is measured by the Euclidean distance between their sparse representations $\vec{t}_i$ and $\vec{t}_j$, namely, $\lVert \vec{t}_i-\vec{t}_j \rVert$.
As a result, $k$-means clustering \citep{lloyd1982} is used to classify $\{\vec{t}_i\}$ into clusters.
A drawback of $k$-means clustering is that the number of clusters is taken as an input parameter. The minimum number of clusters should therefore be determined through extra running of clustering beforehand to make sure the similarities between vectors in the same cluster are acceptable.

Once the vectors $\{\vec{t}_i\}$ are classified into $K$ clusters, a central vector of each cluster is calculated according to the corresponding row vectors $\{\vec{x}_i\}$, as
\begin{equation}
  \vec{h}_k = \frac{1}{N_k} \sum_{i \in S_k} \vec{x}_i \text{,}
\end{equation}
where $N_k$ is the number of vectors in the $k$-th cluster, while $S_k$ is the set of indices of vectors in this cluster.
A circulant matrix $\vec{H}_k$ is then constructed from $\vec{h}_k$ as its first row vector.
Its coefficient matrix $\vec{A}_k$ is constructed from its diagonal elements $\vec{a}_k$, as
\begin{equation}
a_{k,i} = \begin{cases}
1\text{, if } i\in S_k \\
0\text{, otherwise}
\end{cases} \text{.}
\end{equation}
 \section{Test and results}
\subsection{Object, modulation kernels, and observed data}
\begin{figure}[H]
  \includegraphics[width=1.1\linewidth]{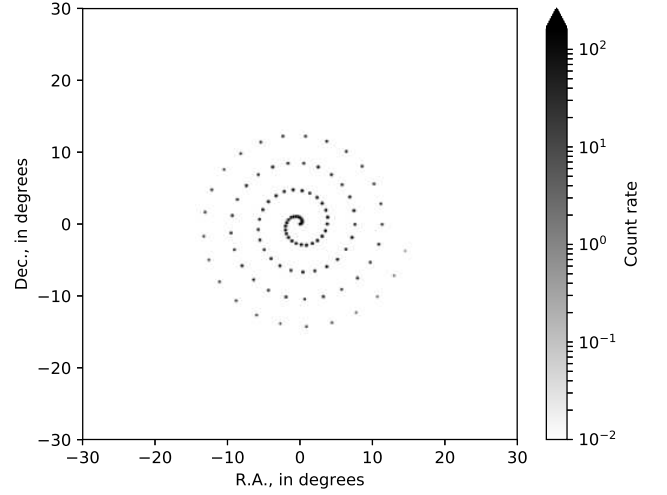}
  \caption{Model object.}
  \label{fig:model}
\end{figure}

We produced a model with a series of point sources located on a spiral as the object (refer to Fig. \ref{fig:model}).
Distance between adjacent sources increases from the interior to the exterior of the spiral, while the intensity decreases.
We can therefore use this model object to assess both spatial resolution and sensitivity of a given observation and the corresponding reconstruction.

\begin{figure}[H]
  \includegraphics[width=1.1\linewidth]{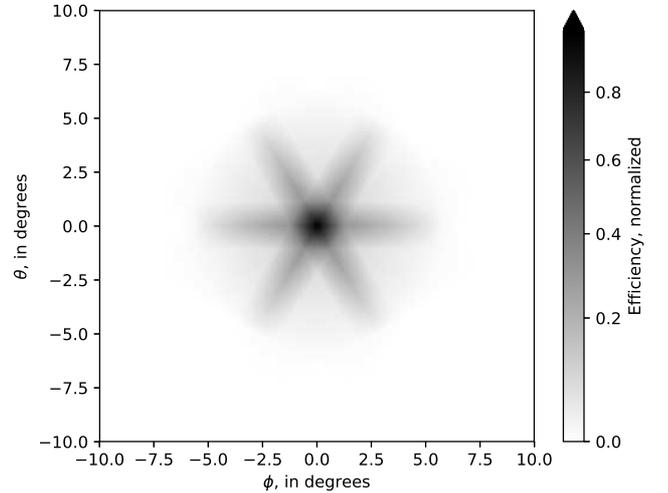}
  \caption{HE/\emph{Insight}-HXMT aperture PSF model, all collimated detectors combined.}
  \label{fig:psf}
\end{figure}

Aperture PSF model of the high energy (HE) telescope of \emph{Insight}-HXMT (as shown in Fig. \ref{fig:psf}), which reflects the geometrical effects of all the collimators of HE/\emph{Insight}-HXMT detectors, that is, the detection efficiency of the telescope to a test point source in its FoV is brought here to build modulation kernels.
The spatial coordinates of the test point source are variables of the PSF.

\begin{figure}[H]
\includegraphics[width=\linewidth]{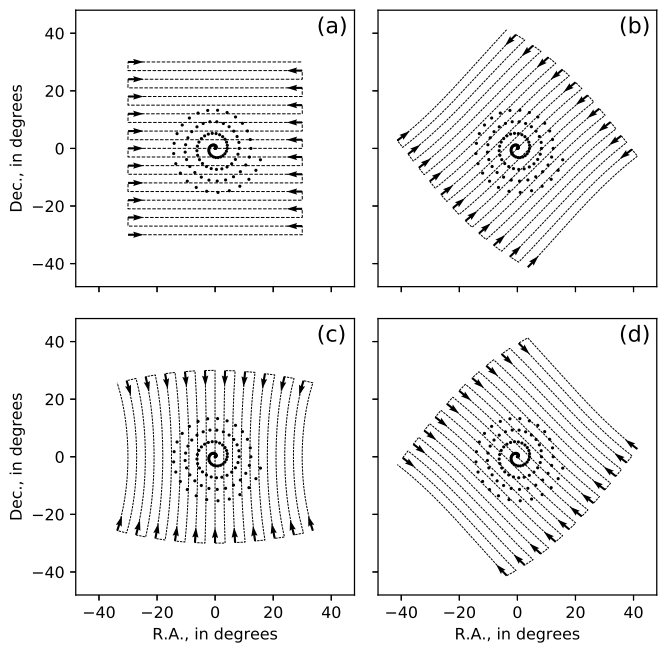}
\caption{Simulated scanning paths to the same object sky region along different directions.}
\label{fig:scan}
\end{figure}

We simulated observations in which the sky region is scanned in different directions as shown in Fig. \ref{fig:scan}.
Scanning speed along each row is $0.10\;^\circ/\mathrm{s}$, while the interval between adjacent scanning rows is $3.0\;^\circ$.
 
\begin{figure}[H]
\includegraphics[width=\linewidth]{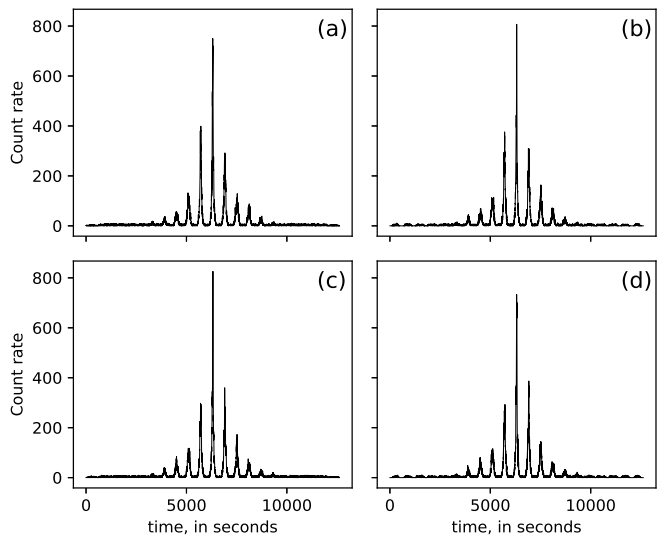}
\caption{Simulated observed light curves in different scanning observations.}
\label{fig:obs}
\end{figure}

With the simulated scanning paths, PSF, object, and a uniform background of $0.3\;\mathrm{counts}/\mathrm{s}/\mathrm{deg}^2$, which is consistent with the scanning paths and rates and the HE background \citep{li2009bgrd}, we calculated the modulation kernel matrices as well as the corresponding modulated light curves, respectively.
As for the detectors, we simulated Poisson fluctuation by generating pseudo Poisson random numbers with the modulated light curves as the expected values.
The simulated observed light curves are shown in Fig. \ref{fig:obs}.
 
\subsection{Reconstruction and comparison}
\begin{figure}[H]
\includegraphics[width=1.1\linewidth]{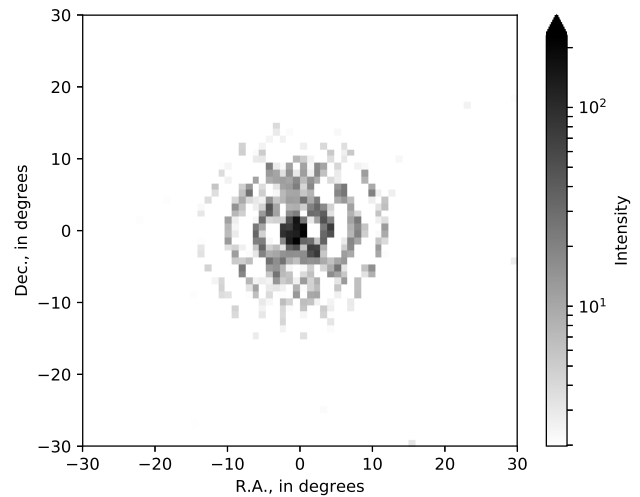}
\caption{Low-resolution image reconstructed from simulated single-epoch light curve.}
\label{fig:recon:lores:single}
\end{figure}

First we reconstructed the object from a single-epoch observed light curve (as shown in the top left panel of Fig. \ref{fig:obs}) on a low-resolution pixel grid (64 by 64 pixels).
The reconstructed image is shown in Fig. \ref{fig:recon:lores:single}.
We estimated from the marginal null region of the reconstructed image that the $3\sigma$ pixel-wise sensitivity is $2.4\;\mathrm{counts}/\mathrm{s}/\mathrm{deg}^2$.
So pixels with image values below the estimated sensitivity were discarded.
 
\begin{figure}[H]
\includegraphics[width=1.1\linewidth]{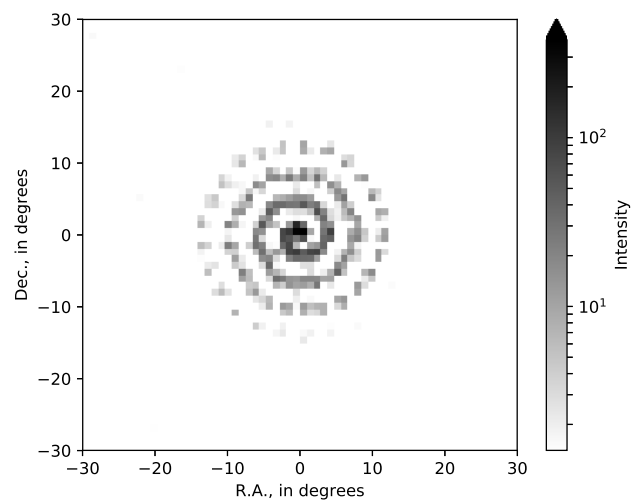}
\caption{Low-resolution image reconstructed from simulated multiple-epoch light curves.}
\label{fig:recon:lores:multiple}
\end{figure}

Similarly we reconstructed the object again from simulated light curves from the four different scanning paths shown in Fig. \ref{fig:scan} by synthesizing all modulation kernel matrices and all light curves together, according to Sect. \ref{sec:multiple-epoch}.
The reconstructed image is shown in Fig. \ref{fig:recon:lores:multiple}.
The estimated $3\sigma$ sensitivity is $1.7\;\mathrm{counts}/\mathrm{s}/\mathrm{deg}^2$.
 
\begin{figure}[H]
  \includegraphics[width=1.1\linewidth]{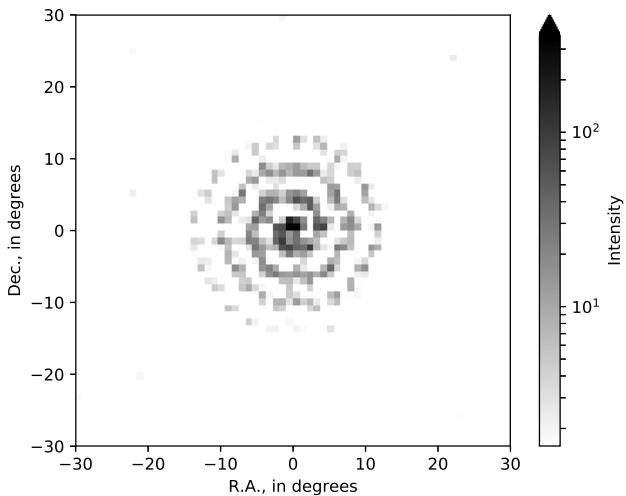}
  \caption{Low-resolution image reconstructed from simulated multiple-epoch light curves with synthetic DD.}
  \label{fig:recon:lores:fft}
\end{figure}

In the above reconstructions we employed the original DD iterations with vector-extrapolation between iterations to speed up the convergence \citep{biggs1997}.
The vector-extrapolation is a technique in numerical analysis which predicts the subsequent step with previous steps instead of computing the iteration,  meaning that less iterations are required.
For the low-resolution reconstruction, we find that after $30$ iterations the solutions no longer improve due to significant artefacts.
For each iteration the modulation was calculated through matrix multiplication routines provided by OpenBLAS, which is a widely used open source implementation of the Basic Linear Algebra Subprograms (BLAS)\citep{openblas}.
This is referred to as the \emph{original DD} while the solution specified in Eq. \ref{equ:syndd} is referred to as the \emph{synthetic DD}.
Time cost per original DD iteration is $5.3$ seconds on a dual-core PC.
We reconstructed the object from simulated light curves from the four different scanning paths shown in Fig. \ref{fig:scan} on the same pixel grid with synthetic DD for comparison, as shown in Fig. \ref{fig:recon:lores:fft}.

Time cost per synthetic DD iteration is $15$ milliseconds on one personal computer (PC).
As we can see the reconstructed image with synthetic DD is similar to that with original DD but the time cost is reduced by orders of magnitude.

\begin{figure}[H]
  \includegraphics[width=1.1\linewidth]{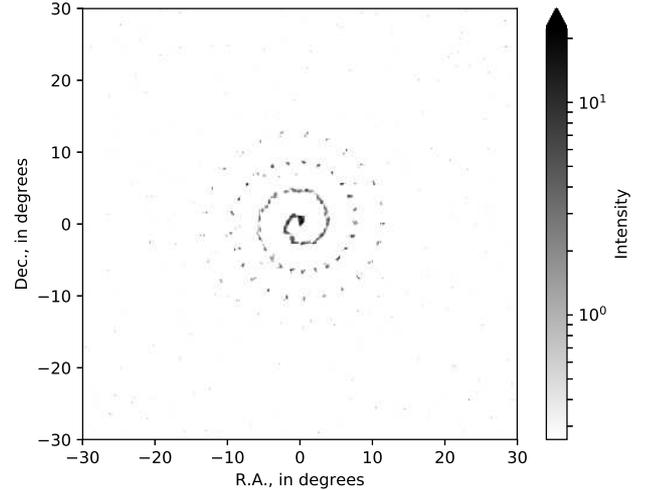}
  \caption{High-resolution image reconstructed from simulated multiple-epoch light curves, with synthetic DD.}
  \label{fig:recon:hires:fft}
\end{figure}

Subsequently, we tested higher-resolution ($512$ by $512$ pixels) reconstruction with synthetic DD.
Original DD was not tested because of its expected time cost (at least $10$ hours).
Time cost per original DD iteration increases polynomially and more iterations are requested with higher resolution.
The image reconstructed from the simulated light curves for the four different scanning paths illustrated in Fig. \ref{fig:scan} is shown in Fig. \ref{fig:recon:hires:fft}.
Time cost per iteration on this pixel grid is $0.6$ seconds.

\begin{figure}[H]
  \includegraphics[width=\linewidth]{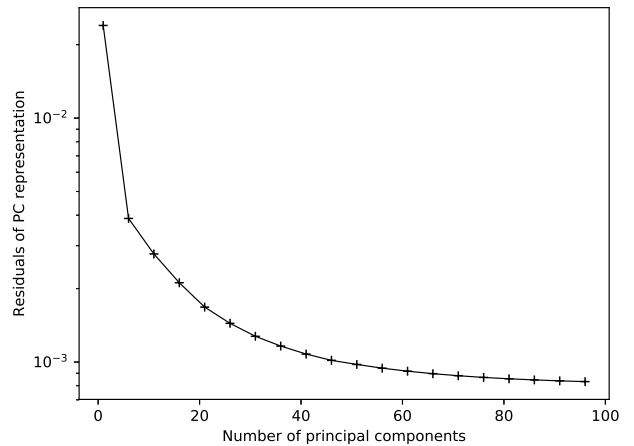}
  \caption{PC representation residuals.}
  \label{fig:pca}
\end{figure}

We also observe the convergence of clustering analysis as a prerequisite of synthetic DD.
The residual of the sparse representation of the kernel matrix ($\frac{\left\vert\vec{H} - \sum_k\vec{A}_k\vec{H}_k\right\vert}{\left\vert\vec{H}\right\vert}$) is reduced rapidly with the first components; subsequently the marginal improvement involving more components appears to diminish, as illustrated in Fig. \ref{fig:pca}.

\begin{figure}[H]
  \includegraphics[width=\linewidth]{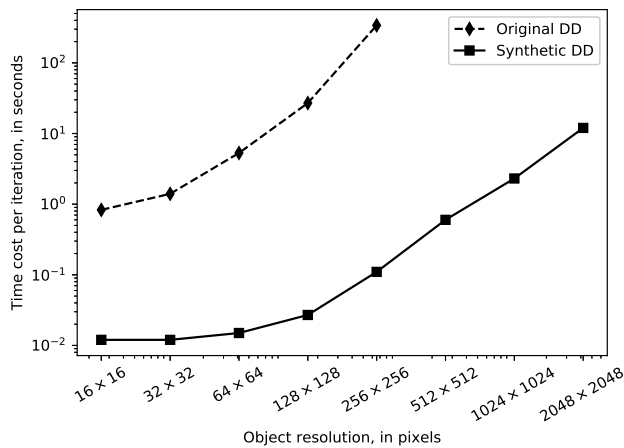}
  \caption{Time costs of reconstructions.}
  \label{fig:timing}
\end{figure}

Finally time costs of reconstructions with original DD iteration and synthetic DD iteration are shown in Fig. \ref{fig:timing}.

\section{Conclusion}
The synDD method including modulation kernel matrix analysis and modulation equation synthesis
reduces the polynomial computational complexity of a time-consuming process in solving the modulation
equation to linear-logarithmic complexity,
which in turn saves time costs of object reconstruction in \emph{Insight}-HXMT data analysis by orders of magnitude.
This encourages us to include more observed data, for example, in multiple epochs or multiple energy bands,
and to reconstruct the object with higher resolution.
Therefore it is possible to achieve both improved sensitivity and higher resolution but with even less
resources.
We suggest using the proposed method in \emph{Insight}-HXMT data analysis especially with scanning observations
from multiple payloads, different energy bands, and/or multiple epochs for transient detection,
time-varying objects monitoring and so on.

\begin{acknowledgements}
Pseudocodes in this article are typeset with the package \textsf{algorithms bundle}.
In this work we made use of SciPy \citep{scipy} and PyTables \citep{pytables} in numerical
computing and dealing with large datasets.
We thank the anonymous referee whose comments and suggestions helped improve and clarify this manuscript.
\end{acknowledgements}


\begin{appendix}
\subsection{Algorithm and codes}
\floatname{algorithm}{Subroutine}
\algsetup{indent=2ex}
The methods we explained above are specified with several separate subroutines in pseudocodes.
In Table \ref{tab:symbols} we have summarised all the symbols and their definitions used in the following subroutines.

Vectors are implemented by 1D arrays.
Matrices are implemented by 2D arrays, where rows and columns are numbered by the first and second indices respectively, that is, $A[i][j]$ represents the matrix element $A_{i,j}$ in the following algorithm descriptions,
$A[i]$ denotes the $i$-th row, and $A[\cdots][j]$ denotes the $j$-th column.
Here ellipsis $\cdots$ between a pair of square brackets $[]$ denotes the sequence of all indices.
Similar notations are used for 3D arrays.
For example, provided that $A$ is an $L \times M \times N$ array of scalar elements, $A[i][j][k]$ represents the element $A_{i,j,k}$.
$A[\cdots][j][k]$, $A[i][\cdots][k]$ and $A[i][j]$ represent vectors.
$A[\cdots][k]$, $A[i][\cdots]$ and $A[\cdots][j][\cdots]$ represent matrices.

\begin{center}
\topcaption{Symbols in algorithm descriptions.} \label{tab:symbols}
\tablefirsthead{\toprule Symbol&\multicolumn{1}{c}{Description} \\ \midrule}
\tablehead{
\multicolumn{2}{c}
{{\bfseries  Continued from previous page}} \\
\toprule
Symbol&\multicolumn{1}{c}{Description}\\ \midrule}
\tabletail{
\midrule \multicolumn{2}{r}{{Continued on next page}} \\ \midrule}
\tablelasttail{
\\\midrule
\multicolumn{2}{r}{{Concluded}} \\ \bottomrule}
\begin{xtabular}{l|p{0.6\linewidth}}
  $\%$                               &   the integer modulo operator \\
$\leftarrow$                       &   the assignment operator \\
$\lVert \cdots \rVert$             &   the Euclidean norm operator \\
$\mathrm{FFT}\left(\cdots\right)$  &   the Fourier transform operator implemented with FFT
                                       algorithm \\
$\mathrm{IFFT}\left(\cdots\right)$ &   the inverse Fourier transform operator \\
$\overline{\cdots}$                &   the element-wise complex conjugation operator \\
$\epsilon$                         &   a small number specified as an input parameter \\
$i$ and $j$                        &   array indices \\
$N$ and $M$                        &   numbers of rows and columns of the original kernel matrix \\
$k$                                &   cluster index \\
$K$                                &   the number of clusters \\
$J$                                &   the expected number of principal components \\
$d$                                &   array of $N$ elements represents the observed data \\
$H$                                &   $N \times M$ array representing the original kernel matrix \\
$X$                                &   $N \times M$ array representing the shifted kernel matrix so that row vectors are aligned \\
$X_0$                              &   array of $N$ elements storing the arithmetic average of the
                                       $M$ column vectors of $X$ \\
$T$                                &   $N \times J$ array representing the principal component scores of the
                                       $N$ row vectors of $X$, i.e., $T[i]$ is the principal component score
                                       of $X[i]$ \\
$R$                                &   $N \times M$ array storing residuals in PCA Subroutine
                                       \ref{sub:pca} \\
$\lambda$ and ${\lambda}_\text{p}$  &   estimates of the eigenvalue of the covariance
                                       in Subroutine \ref{sub:pca} \\
$S$                                &   array of $K$ lists, while the $k$-th list $S[k]$ contains all the
                                       row numbers of vectors classified into the $k$-th cluster at the
                                       current step \\
$L$                                &   array of $K$ elements representing the numbers of vectors in the $K$
                                       clusters, e.g. $L[k]$ is the number of vectors in the $k$-th
                                       cluster \\
$C$ (and $C_\text{p}$)              &   $K \times J$ array, the $k$-th row of which stores the central
                                       vector of the $k$-th cluster calculated at the current step
                                       (and the previous step),
                                       represented by principal component scores \\
$Y$                                &   an $K \times M$ array storing the central vectors of each cluster in
                                       pixel-wise representation \\
$D$                                &   an $K \times N$ array representing the Euclidean distance between the $N$
                                       shifted vectors and central vectors of the $K$ clusters, e.g.
                                       $D[k][i]$ represents the Euclidean distance between the $i$-th shifted
                                       vector $X[i]$ and the $k$-th central vector $C[k]$ \\
$P$                                &   an $K \times N \times M$ array representing the $K$ circulant kernel
                                       matrices in Eq. \ref{equ:decompose} \\
$A$                                &   an $K \times N \times N$ array representing the $K$ diagonal coefficient
                                       matrices in Eq. \ref{equ:decompose} \\
$a$                                &   an $K \times N$ array representing the vectors of diagonal elements of
                                       the $K$ matrices \\
$w$                                &   an array of $M$ elements representing the normalization factor in Eq.
                                       \ref{equ:syndd} \\
$f$ (and $f_\text{p}$)              &   array of $M$ elements representing the object image estimated at the
                                       current step (and the previous step) \\
$r$                                &   an array of $N$ elements \\
$q$                                &   an array of $M$ elements \\
 \end{xtabular}
\end{center}

The circularly-shifting operation mentioned in Sect. \ref{sec:cluster} is specified with Subroutine \ref{sub:cshift}, which is usually implemented by built-in functions in numerical computing languages such as MATLAB, NumPy, and so on.

\begin{algorithm}[H]
\begin{algorithmic}
\REQUIRE $M,N,H$
\FOR{$i=1$ \TO $N$}
  \FOR{$j=1$ \TO $M$}
    \STATE $X[i][j] \leftarrow H[i][(j+i-2) \% M + 1]$
  \ENDFOR
\ENDFOR
\RETURN $X$
\end{algorithmic}
 \caption{Circularly-shift row vectors.}
\label{sub:cshift}
\end{algorithm}

A resumable implementation of the NIPALS method for PCA is specified in Subroutine \ref{sub:pca}.
If the principal component scores matrix $T$ is not initialized for writing, for example, when no memory or file system resources have been allocated to it, the current call to this subroutine will be considered as the first run.
In this case the necessary resources will be allocated to the score matrix $T$, the loading matrix $P$ as well as the residual $R$ inside the subroutine during the current call.
Otherwise it will be considered as resumed from a previous run, where the resources that $T$, $P$ and $R$ refer to must be preserved.
The expected number of principal components $J$ is provided as an input parameter.
Elements of each principal component score are computed iteratively.
An iteration is stopped when the increment of the estimate of the eigenvalue becomes negligible \citep{geladi1986}.

\begin{algorithm}[H]
\begin{algorithmic}
\REQUIRE $M,N,J,X,R,P,T,\epsilon$
\STATE $j \leftarrow 1$
\IF{resumed from previous run}
    \WHILE{$j \leq J$ and $\lVert T[\cdots][j] \rVert \leq\epsilon$}
        \STATE $j \leftarrow j+1$
    \ENDWHILE
\ELSE[this is the first run]
    \FOR{$i=1$ \TO $N$}
        \STATE $X_0[i] \leftarrow 0$
    \ENDFOR
    \FOR{$i=1$ \TO $M$}
        \STATE $X_0 \leftarrow X_0 + X[\cdots][i]$
    \ENDFOR
    \STATE $X_0 \leftarrow \sfrac{X_0}{M}$
    \FOR{$i=1$ \TO $M$}
        \STATE $R[\cdots][i] \leftarrow X[\cdots][i]-X_0$
    \ENDFOR
\ENDIF
\IF[all expected PC scores have been calculated]{$j > J$}
    \RETURN $R,P,T$
\ENDIF
\FOR{$j=j$ to $J$}
    \STATE $\lambda \leftarrow 0$
    \STATE $T[\cdots][j] \leftarrow R[\cdots][j]$
    \REPEAT
        \FOR{$i=1$ \TO $M$}
            \STATE $P[i][j] \leftarrow$ sum of $R[\cdots][i] \cdot T[\cdots][j]$
        \ENDFOR
        \STATE $P[\cdots][j]\leftarrow \sfrac{P[\cdots][j]}{\lVert P[\cdots][j] \rVert}$
        \FOR{$i=1$ \TO $N$}
            \STATE $T[i][j] \leftarrow$ sum of $R[i] \cdot P[\cdots][j]$
        \ENDFOR
        \STATE ${\lambda}_\text{p} \leftarrow \lambda$
        \STATE $\lambda \leftarrow \lVert T[\cdots][j]\rVert$
    \UNTIL{$\lVert {\lambda}_\text{p} - \lambda \rVert \leq \sfrac{\epsilon}{2} \cdot \lVert {\lambda}_\text{p} + \lambda \rVert$}
    \FOR{$i=1$ \TO $M$}
        \STATE $R[\cdots][i] \leftarrow R[\cdots][i] - T[\cdots][j] \cdot P[i][j]$
    \ENDFOR
\ENDFOR
\RETURN $R,P,T$
\end{algorithmic}
 \caption{NIPALS-PCA}
\label{sub:pca}
\end{algorithm}

The $k$-means clustering is specified with Subroutine \ref{sub:clustering}.
We randomly choose $K$ vectors as the initial central vectors of the $K$ clusters.
The loop is stopped when none of the central vectors changes any more.

\begin{algorithm}[H]
\begin{algorithmic}
\REQUIRE $K,N,T,X,\epsilon$
\FOR{$k=1$ \TO $K$}
  \STATE $i \leftarrow \text{a random integer from $1$ to $N$}$
  \STATE $C[k] \leftarrow T[i]$
  \STATE $Y[k] \leftarrow 0$
\ENDFOR
\STATE $\delta \leftarrow 0$
\REPEAT
  \FOR{$i=1$ \TO $N$}
    \FOR{$k=1$ \TO $K$}
      \STATE $D[k][i] \leftarrow \lVert T[i]-C[k] \rVert$
    \ENDFOR
    \STATE $k \leftarrow$ index of the least element of $D[\cdots][i]$
    \STATE append $i$ to $S[k]$
  \ENDFOR
  \FOR{$k=1$ \TO $K$}
    \STATE $L[k] \leftarrow$ number of elements in $S[k]$
    \STATE $C_\text{p}[k] \leftarrow C[k]$
    \STATE $C[k] \leftarrow 0$
    \FOR{$i=1$ \TO $L[k]$}
      \STATE $C[k] \leftarrow C[k]+T[S[i]]$
    \ENDFOR
    \STATE $C[k] \leftarrow \sfrac{C[k]}{L[k]}$
    \STATE $\delta \leftarrow \max[\delta, \lVert C_\text{p}[k] - C[k] \rVert ]$
  \ENDFOR
\UNTIL{$\delta \leq \epsilon$}
\FOR{$k=1$ \TO $K$}
  \FOR{$i=1$ \TO $L[k]$}
    \STATE $Y[k] \leftarrow Y[k]+X[S[i]]$
  \ENDFOR
  \STATE $Y[k] \leftarrow \sfrac{Y[k]}{L[k]}$
\ENDFOR
\RETURN $A,C,Y$
\end{algorithmic}
 \caption{$k$-means clustering on sparse representation.}
\label{sub:clustering}
\end{algorithm}

The iterative solution of synthetic modulation equation formulated in Eq. \ref{equ:syndd} is specified in Subroutine \ref{sub:mliter}.
If $M=N$ the observed data $d$ itself can serve as the initial estimate, otherwise its one-fold correlation $\vec{H}^\mathrm{T}\vec{d}$ is provided as initial estimate of the object.
Convolutions in Eq. \ref{equ:syndd} are implemented with FFTs.
The iteration is stopped when the difference between the current estimate of the image and the previous one is negligible.
The convergence of the iteration can be accelerated via vector extrapolation \citep{biggs1997}.

\begin{algorithm}[H]
\begin{algorithmic}
\REQUIRE $K,M,N,A,Y$
\STATE $f \leftarrow$ initial estimate
\FOR{$k=1$ \TO $K$}
  \FOR{$i=1$ \TO $N$}
    \STATE $a[k] \leftarrow A[k][i][i]$
  \ENDFOR
\ENDFOR
\FOR{$i=1$ \TO $M$}
  \STATE $w[i] \leftarrow 0$
\ENDFOR
\FOR{$k=1$ \TO $K$}
  \STATE $w \leftarrow w+\mathrm{IFFT}\left(\mathrm{FFT}\left(a[k]\right)\cdot\mathrm{FFT}\left(Y[k]\right)\right)$
\ENDFOR
\REPEAT
  \STATE $f_\text{p} \leftarrow f$
  \FOR{$i=1$ \TO $M$}
    \STATE $q[i] \leftarrow 0$
  \ENDFOR
  \FOR{$i=1$ \TO $N$}
    \STATE $r[i] \leftarrow 0$
  \ENDFOR
  \FOR{$k=1$ \TO $K$}
    \STATE $r \leftarrow r + a[k]\cdot \mathrm{IFFT}\left( \mathrm{FFT}\left(f_\text{p}\right) \cdot \overline{ \mathrm{FFT}\left( Y[k] \right)} \right)$
  \ENDFOR
  \STATE $r \leftarrow \sfrac{d}{r}$
  \FOR{$k=1$ \TO $K$}
    \STATE $q \leftarrow q + a[k]\cdot \mathrm{IFFT}\left( \mathrm{FFT}\left(r\right) \cdot \mathrm{FFT}\left(Y[k]\right) \right)$
  \ENDFOR
  \STATE $f \leftarrow \sfrac{f_\text{p} \cdot q}{w}$
\UNTIL{$\lVert f-f\text{p}\rVert \leq \epsilon$}
\RETURN $f$
\end{algorithmic}
 \caption{Iterative ML demodulation.}
\label{sub:mliter}
\end{algorithm}

\begin{figure}[H]
\centering
\includegraphics[width=\linewidth]{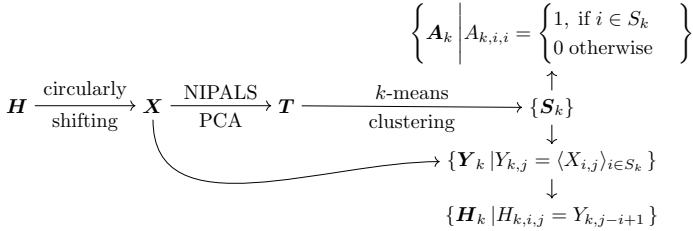}
\caption{Algorithms used to decompose a kernel matrix.}
\label{fig:decompose}
\end{figure}

The workflow of algorithms used to decompose a kernel matrix $\vec{H}$ is summarised in Fig. \ref{fig:decompose}.
 \end{appendix}

\end{document}